\documentclass{iaus}
\usepackage{graphicx}
\usepackage[a4paper,breaklinks,dvipdfm]{hyperref}
\newcommand{\mlp}{\ensuremath{\alpha_{\mathrm{MLT}}}}
\newcommand{\cobold}{\mbox{CO$^5$BOLD}}
\newcommand{\Teff}{\mbox{$T_{\mathrm{eff}}$}}
\newcommand{\logg}{\mbox{$\log g$}}

\title[LTE model atmospheres  ] 
{LTE model atmopsheres \\ MARCS, ATLAS and CO5BOLD}

\author[Bonifacio, Caffau, Ludwig \& Steffen]   
{P. Bonifacio$^1$
 \and E. Caffau$^{2,1}$
\and H.-G. Ludwig$^{2,1}$
\and M. Steffen$^{3,1}$
}

\affiliation{$^1$
GEPI, Observatoire de Paris, CNRS, Universit\'e Paris Diderot\\
Place Jules Janssen, 92190 Meudon, France
\\[\affilskip]
$^2$
Zentrum f\"ur Astronomie der Universit\"at Heidelberg, Landessternwarte\\
K\"onigstuhl 12, 69117 Heidelberg, Germany\\[\affilskip]
$^3$
Leibniz-Institut f\"ur Astrophysik Potsdam\\
An der Sternwarte 16, D-14482 Potsdam, Germany
}

\pubyear{2011}
\volume{282}  
\pagerange{110--117}
\jname{From Interacting Binaries to Exoplanets; Essential modelling
Tools}
\editors{M. Richards ed.}
\begin{document}

\maketitle

\begin{abstract}
In this talk we review the basic assumptions and
physics covered by classical 1D LTE model atmospheres.
We will focus on ATLAS and MARCS models of F-G-K stars 
and describe what resources are available through the web, 
both in terms of codes and model-atmosphere grids. 
We describe the 
advances made in hydrodynamical simulations of convective stellar
atmospheres with the \cobold\ code
 and what
grids and resources are available, with a prospect of what
will be available in the near future.  
\end{abstract}

\firstsection 
\section{Introduction}

A model atmosphere is a numerical model that 
describes the physical state of the plasma in the outer layers of a star, 
and is
used to compute observable quantities, such as the emerging
spectrum or colours.
Different degrees of complexity lead to different classes of models.
The first simplification that is made in the models
we shall describe is that of Local Thermodynamic Equilibrium (LTE).
Although we know that a stellar atmosphere cannot be in
thermodynamic equilibrium, since we see that radiation is escaping
in open space, we make the assumption that {\em locally}
we are very close to thermodynamic equilibrium. 
In practice this means that we assume that at each point in 
the atmosphere, which we identify by some
suitable coordinates $X,Y,Z$ and in a volume $\Delta X \Delta Y \Delta Z$
around it, we can define a temperature T.
This temperature can be used to describe the velocity distribution
of the particles, that will be a Maxwellian distribution, to compute
the occupation numbers of the atomic levels of the different
species, through Boltzmann's law and the ionisation equilibria
through Saha's law.
The LTE hypothesis greately simplifies the computation
of model atmospheres, by providing us
a quick and easy way to compute all the micro-physics of the 
plasma, just by knowing the temperature (and gas pressure)
at any given point.

The next simplification  concerns the
dimensionality of the problem. Do we really need to treat
this as a three-dimensional problem ?
If we simplify our model by assuming that the atmosphere is 
homogeneous in two directions and shows variations
of the physical quantities only along one direction
(vertical, for plane-parallal atmosheres, radial for
spherical atmospheres), we have reduced our problem
to  one dimension.

In principle all the physical quantities in the plasma
may change with time. A simplification is to assume
that the atmosphere is stationary and then the problem
becomes time-independent. 

All these simplifications are adopted in the widely used
MARCS and ATLAS model atmospheres. In spite of what 
can appear at first sight as an oversimplification, 
these model atmospheres are capable of reproducing
a wide range of observable quantities and have a
high predictive power.

The \cobold\ model atmospheres make the hypothesis
of LTE, however are fully three dimensional (two dimensional
models may also be computed) and time-dependent. In this
respect they are more realistic, since they can describe
effects that cannot be accounted for by 1D models.
This extra realism, however comes at a price and we shall
discuss this later.

In the following we shall give only a very sketchy 
description of the basic principles that are at 
the basis for the computation of model atmospheres,
since these are well described in the relevant publications. 
We shall instead try to adopt an "end user" approach
pointing out how such tools can be used and what resources
are available.

\section{ATLAS and MARCS}

\subsection{Basic principles}

Both the ATLAS \cite[(Kurucz 1970,1993,2005)]{K70,K93,K05}
and MARCS 
 \cite[(Gustafson et al. 1975; Plez et al. 1992; Edvardssoon et al. 1993; Asplund et al. 1997; Gustafsson et al. 2003,2008)]{gustafsson75,plez92,edvardsson93,asplund97,gustafsson03,gustafsson08}
models are one dimensional
and static. MARCS can deal either with plane-parallel 
or spherical geometry, ATLAS only
with plane-parallel, although SATLAS \cite[(Lester \& Neilson 2008)]{lester} 
can
compute spherical models. Elswhere in this
volume Neilson \cite[(2011)]{Neilson} talks about SATLAS
and spherical models. In what follows we shall
assume plane-parallel geometry for both MARCS
and ATLAS.

Both codes assume that the atmosphere is in hydrostatic
equilibrium, this provides the first basic equation necessary
to compute a model  atmosphere.
The equation of hydrostatic  equilibrium 
states that the gas pressure gradient is balanced by the difference between
gravity and the sum of turbulent pressure gradient and radiative acceleration.
In MARCS the turbulent pressure gradient is treated
defining an ``effective gravity'' $g_{\rm eff}$, see 
\cite[Gustafsson et al. (2008)]{gustfsson08}. 
In ATLAS there is an explicit term for the turbulent
pressure, 
proportional to density and the square
of turbulent velocity.
The equation 
of hydrostatic equilibrium
must be coupled with the equation of radiative transfer, 
and the condition of energy conservation. The atmosphere simply transports
the  energy, there is no net absorption or creation of energy 
within the atmosphere.
A convenient variable is the mass column density
$
dm = -\rho dx
$, RHOX (pronounced ``rocks'' for those who like reading
the {\tt FORTRAN}).
The goal of the model atmosphere computation
is to define the run of the temperature $T(m)$, and the energy
conservation provides a mean to modify a trial value of $T$ in order
to satisfy the condition.
The global parameters that define the model are the surface gravity $g$,
the energy/surface = integral of flux over all frequencies = 
$\sigma\rm T_{eff}^4$, where the latter can be taken as a definition of
``effective temperature'' and the ``chemical composition'', that
affects the opacities and the equation of state.
At each step in the computation the hypothesis of LTE allows
to compute occupation numbers and ionization fractions that allow
to compute the opacity.
ATLAS has two ways of dealing with line opacities, either through
the Opacity Distribution Functions (ODFs, version 9 of ATLAS)
or through Opacity Sampling (version 12 of ATLAS).
Early versions of MARCS also used ODFs, but the currently used version
is OSMARCS, that uses opacity sampling.
We stress that tests conducted with ATLAS 9 and ATLAS 12
show that the differences between a model computed with ODFs and one
computed with Opacity Sampling are minor and can be ignored for
all practical purposes. The choice on whether to use ATLAS 9
or ATLAS 12 is a matter of convenience, if a large number of models
needs to be computed with the same chemical composition, then 
ATLAS 9 is the obvious choice. If several models with slightly 
different chemical composition are required, ATLAS 12 is more handy.

In the atmospheres of cool stars, a complication is that in the deep layers 
energy is mainly transported by convection. 
Both MARCS and ATLAS use a ``mixing length'' approximation. 
However MARCS uses the \cite[Henyey et al. (1965)]{henyey}
formulation, while ATLAS uses essentially the
\cite[Mihalas (1970)]{mihalas} formulation, in a way
that is detailed by \cite[Castelli, Gratton \& Kurucz (1997)]{CGK}. 
At large optical depths where convection dominates,
a consequence is that a MARCS and an ATLAS model will be slightly different, 
even if they have been computed
with the same mixing length parameter \mlp.
The effect is illustrated in \cite[Bonifacio et al. (2009)]{B09},
appendix A, figure A.2.
We wish to give here a warning: ATLAS has an option
for an ``approximate treatment of overshooting'', well
described by \cite[Castelli et al. (1997)]{CGK}, that 
we recommend users to switch off. This is 
also the recommendation of \cite[Castelli et al. (1997)]{CGK},
but in our view the most convincing reason is that
this option produces a temperature structure that is
inconsistent with the mean temperature structure of
hydrodynamical simulations.

\subsection{Availability}

The ATLAS code has always been distributed publicly, on an
``as is'' basis and a ``do not use blindly'' clause.
The main site is Kurucz's site
\url{kurucz.harvard.edu} were you can find all the source
codes of versions 9 and 12 of ATLAS, as well as the 
spectrum synthesis suite SYNTHE, the abundance analisis
code WIDTH and a lot more.
We call attention to a code called {\tt binary} that allows to 
combine two synthetic spectra to sythesise the
spectrum of an SB2 binary.
The site also contains a large choice of ODFs, atomic data
to compute new ODFs, atomic and molecular data for spectrum
synthesis. There is also a large grid of computed ATLAS 9 models, 
that can be used ``off the shelf''. 

One difficulty faced by ATLAS users is that Kurucz uses 
DEC computers running under the VMS operating system.
While such systems were widely spread in the
eighties, and still existent in the ninties, the majority
of researchers has access to Unix work stations, and a large
fraction are Linux systems. 
\cite[Sbordone et al. (2004)]{SB04} presented a port of ATLAS and
the other codes for Linux, see also \cite[Sbordone (2005)]{SB05}.

Fiorella Castelli is very active in updating and improving
the codes, the latest version of the Linux version of the codes
can always be found on her web site 
\hyperref[wwwuser.oats.inaf.it/castelli]{wwwuser.oats.inaf.it/castelli}.
On the site you can also find a large grid of computed ATLAS 9 models
and fluxes, as well as ODFs to compute further models you might need.

With Luca Sbordone and Fiorella Castelli we also provided a
site were the codes are nicely bound in tar-balls and come
with a  {\tt Makefile} that allows an easy installation. 
The site attempts also to collect available documentation and
example scripts to run the various programmes, to provide a starting
  point for beginners. The site was initially hosted by
the Trieste Observatory, but has now moved to the
Paris Observatory \url{atmos.obspm.fr}.
We try to keep the source codes always aligned with those
on the site of Fiorella Castelli.

For users of ATLAS and related codes there are discussion
and announcement lists that are mantained at the
University of Ljubljana:\\
\url{list.fmf.uni-lj.si/mailman/listinfo/kurucz-discuss}.

The MARCS code is not publicly available, so if you 
need a particular MARCS model you have to ask one of its developers.
However there is a web site on which a large grid
of computed models and fluxes is publicly available:
\url{marcs.astro.uu.se}.
You must register to the site, but registration is free.
You find both plane-parallel and spherical models,
as well as programs to read the models and interpolate
in the grid.
If you use a spherical model, make sure that the
spectrum synthesis code you use is capable of properly
treating the spherical transfer.
For example SYNTHE is not capable of doing it, it will 
nevertheless run happily interpeting the spherical
model as if it were plane-paralle, which is inconsistent.
The {\tt turbospectrum} code by B. Plez \cite[Alvarez \& Plez (1998)]{turbo}
is capable of treating correctly both spherical and plane-parallel models.

Both ATLAS and MARCS are ``state of the art'' 1D model atmosheres. 
In the range of F-G-K stars the differences between the two
kinds of models are immaterial, as shown e.g. in 
\cite[Bonifacio et al. (2009)]{B09}. For the very cool models
(below 3750 K) MARCS models are probably more reliable, 
because all the relevant molecular opacities are included.
ATLAS can compute models also for A-B-O stars, although 
for stars hotter than 20\,000\,K the hypothesis of LTE
clearly becomes questionable. MARCS only computes models
up to 8000 K, therefore if you are dealing with A-F stars, 
ATLAS is preferable.

\section{CO$^5$BOLD models}

\begin{figure}
\begin{center}
\resizebox{0.95\hsize}{!}{\includegraphics[clip=true,angle=90]{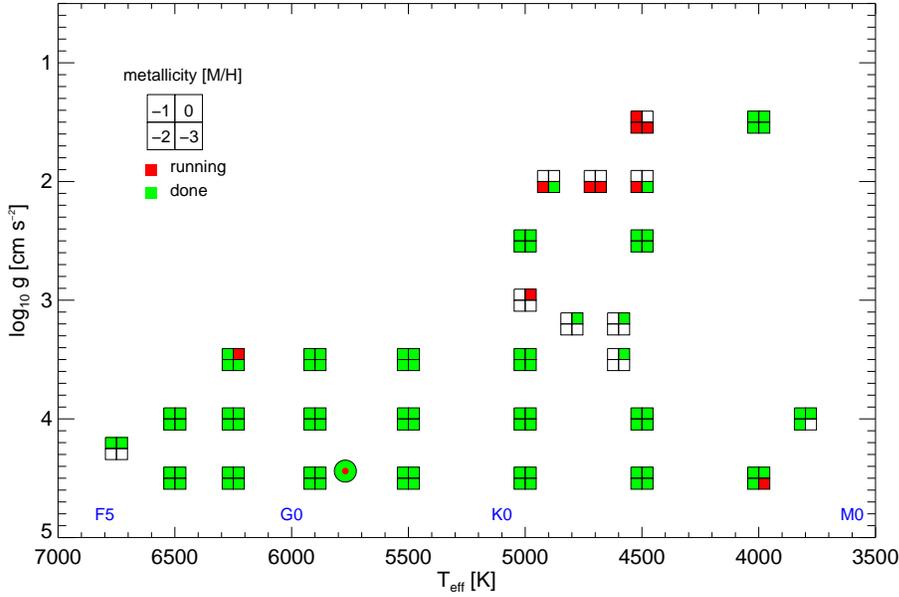} }
 \caption{The current status of the CIFIST grid.
This is an update of the figure shown by 
\cite[Ludwig et al. (2009)]{cifist}
  Symbols mark the location of a model in the \Teff-\logg-plane. Green (gray)
  color indicates completed model runs, red (black) ongoing calculations.
  Each square is split into four sub-squares indicating solar, 1/10, 1/100,
  and 1/1000 of solar metallicity. The solar position is indicated by the
  round symbol. In addition to the models shown in this
plot there are a few low gravity models that are beeing
computed in Vilnius by Arunas  Ku{\v c}inskas
and his collaborators.}
   \label{cifist}
\end{center}
\end{figure}

\cobold\ stands for COnservative COde for the COmputation
of COmpressible COnvection in a BOx of L Dimensions with l=2,3
and is developed by B. Freytag  and M. Steffen with contributions of
H.-G. Ludwig, W. Schaffenberger, O. Steiner, and S. Wedemeyer-B\"ohm
\cite[(Freytag et al. 2002, 2011)]{freytag02,freytag10}.

What it does is to solve the time-dependent equations of
compressible hydrodynamics in an external gravity field
coupled with the non-local frequency-dependent radiation transport. 
It can operate in two modes: 1) the ``box-in-a-star'' mode, 
in which the computational domain covers a small
portion of the stellar surface; 2) the ``star-in-a-box'' mode, 
in which the computational domain includes the whole convective envelope
  of a star.

The need to go to the hydrodynamical  model is obvious: all
the phenomena that are linked to the motions in the
stellar atmospheres, like line-shifts, line asymmetries,
micro-variability etc. cannot be described by 1D static models
like MARCS or ATLAS.

There are  cases in which the use of 1D models
will provide the wrong results, thus 
making the use of 3D models 
mandatory.
One case is the measurement of $^6$Li 
in metal-poor warm dwarf stars.
\cite[Cayrel et al. (2007)]{cayrel07}
have shown that the line asymmetry due to convection
can mimic the presence of up to 5\% $^6$Li,
thus an analysis of the spectra based on 1D model atmospheres,
that provide only symmetric line profiles, will result
in a $^6$Li abundance that is spurious, even if no $^6$Li
is present. In fact this is a misinterpretation of
the convective line asymmetry.
Another case is the measurement of the abundance
of thorium in the solar photosphere. The only line 
suitable for this purpose is the 401.9\,nm Th{\sc ii} resonance
line, that
lies on the red wing of a strong blend of Ni and Fe. 
\cite[Caffau et al. (2008)]{caffau08} have shown
that neglect of the line asymmetry of this Ni-Fe blend
would result in an over-estimate of the Th abundance
by about 0.1\,dex.
Another effect that is very important in metal-poor
stars is the so called ``overcooling''
\cite[(Asplund et al. 1999; Collet et al. 2007;
Caffau \& Ludwig 2007; 
Gonz\'alez Hern\'andez et al. 2008;
Bonifacio et al. 2010)]{asp99,collet2007,CL07,jonay,B10}.
The hydrodynamical models predict much cooler temperatures
in the outer layers than do the 1D models,  resulting
in important differences in the computed line strength
for all the lines that form in these outer layers. 
This is the case for all the molecular species usually
used for abundance determinations \cite[(Behara et al. 2009)]{behara},
but also for some atomic species \cite[(Bonifacio et al. 2010)]{B10}.

The larger amount of information provided by the 3D models comes
at a larger computational cost and the need
to simplify the treatment of some physical effects.
The first necessary simplification is the treatment
of opacity. In the computation of an hydrodynamical model
we cannot afford a treatment of opacity as detailed as
we can in 1D. For this reason the opacities
are grouped into a small number (less or equal to 14)
of opacity bins 
\cite[(Nordlund 1982; Ludwig 1992; Ludwig et al. 1994;
Voegler et al. 2004)]{Nordlund82,Ludwig92,Ludwig+al94,Voegler+al04}.
Further simplifications are currently an approximate treatment of
  scattering in the continuum, and neglecting effects of line shifts in the
  evaluation of the line blocking.

\subsection{The CIFIST grid}

A hydrodynamical model is not easy to perform since it can need several
  months for dwarf stars up to more than a year for giants (on PC-like
    machines). Typically, one ends-up with a time series covered by about 100
    snapshots representing a couple of convective turn-over times.  These
  occupy several GB of storage space.  The computational and human effort to
  compute an hydrodynamical model is such that we cannot expect to be able
to  compute a model for any set of input parameters on a 
working time scale of
  few weeks.

Furthermore, not all snapshots are statistically independent.
It would thus be  not wise to attempt a ``brute force'' approach
and compute the emerging spectrum from each and every snapshot.
A preferable strategy is to select a sub-sample of the
snapshots that has  the same global statistical properties
as the total ensemble \cite[(Caffau 2009)]{tesi}.  
However, even selecting only 15 to 20 snapshots, the 
line formation computations for a large number of lines, 
as are usually employed in the analysis with 1D
models, is computationally demanding.

For these reasons in the course of the CIFIST project
(Cosmological Impact of the FIrst STars, 
\url{cifist.obspm.fr}) we decided to attempt the computation
of a complete grid of hydrodynamical models. 
The foreseen use of this grid is that obseravable quantities
are computed on the grid, so that they can then be conveniently
interpolated for any value within the grid points.
An example of this is provided in \cite[Sbordone et al. (2010)]{S10}
where a fitting function is provided to the curves of growth
of the Li{\sc i} doublet computed on the grid.
One inputs the measured equivalent width, effective temperature, 
surface gravity and metallicity and the fitting function
provides the lithium abundance. 
Another example is \cite[Gonz\`alez Hern\`andez et al. (2010)]{JGH}
where the OH lines have been computed on the grid and the results
may be used for abundance analysis.

Such ready-to-use tools are probably more useful
to researchers than providing the 3D models ``as is''.
We are thus concentrating on computing a meaningful
set of observable quantities so that the \cobold\
models can be widely used for abundance analysis.

\begin{discussion}

\discuss{Tout}
{Can you not get around the time taken to complete the full 3-D 
calculation by something like a 
two-stream model with large slowly 
rising cells and smaller rapidly falling cells?}

\discuss{Bonifacio}
{No. This was the approach tried by R.L. Kurucz in ATLAS 11,
a version of ATLAS that was never released for public use, 
and employed by him  to assess the effect of granulation on the line formation
of the lithium doublet in Pop II stars (Kurucz 1995, ApJ 452, 102).
His result was that the Li abundance should be higher 
than what derived from an analysis with ATLAS 9 by almost one order
of magnitude. This result is totally wrong. We now know that
when treated in LTE the effect of granulation is to {\em lower}
the Li abundance by 0.2 to 0.3 \, dex (Asplund et al. 1999 A\&A 346 L17).
The  two stream model was certainly worth exploring,
at the time it looked to me a brilliant idea.
In retrospect we can see why it failed: to compute each stream in
a 1D approach you have to assume energy conservation, {\em for each stream}.
This condition is clearly violated  and with this approach you cannot take
into account the energy exchanges between the streams.}

\discuss{Petr Harmanec}
{After listening to your excellent talk, 
I am about to commit a suicide. 
Do you think that the 3D model atmosphere models will 
get to the level that non-experts 
would be able to use them for a reliable comparison with real observations?}

\discuss{Bonifacio}
{Indeed we  hope 3D models may become
of general use, we think that the most promising 
approach is to have a grid of 3D model atmospheres,
like the CIFIST grid, and provide to users observable
quantities computed across the grid and means to interpolate
in between grid points. A good example is the
fitting function for the Li abundance provided by 
Sbordone et al. (2010).
I stress again that in the computation of  3D
model atmospheres and associated line formation
we make some simplifications. Notably 
line opacity and  scattering  are treated in
more detail in a 1D computation.}

\discuss{A. Pr\v{s}a}
{Your 3D models are computed for spherical stars. 
If one wanted to compute a spectrum of a distorted star, 
could one compute the intensities for each 
tile of the discretized surface and then sum them up?
}
\discuss{Bonifacio}
{The spherical geometry is dealt with in the star-in-a-box
models. It looks to me more correct to treat distorsion
to sphericity directly in this kind of computation, rather
than tiling several box-in-a-star models.
This has never been done so far, to my knowledge.
I suggest you contact B. Freytag, who has been computing
many star-in-a-box \cobold\ models and can answer
your question more fully than myself.
I am not a \cobold\ developer, I am just an end user.}

\end{discussion}

\end{document}